\documentclass[aps, prb, preprint, lengthcheck, footinbib, showpacs, superscriptaddress, citeautoscript]{revtex4-1}
\usepackage[T1]{fontenc}
\usepackage[english]{babel}
\usepackage{amsmath}
\usepackage{braket}
\usepackage{graphicx}
\usepackage[colorlinks, urlcolor=blue, citecolor=blue, linkcolor=blue, pdfstartview=FitH]{hyperref}
\usepackage[all]{hypcap}

\begin{document}

\def\affiSOLAB{Spin Optics Laboratory, Saint~Petersburg State University, 198504 St.~Petersburg, Russia}
\def\affiRC{SPbU Resource Center ``Nanophotonics'', Saint~Petersburg State University, 198504 St.~Petersburg, Russia}
\def\affiSP{Departament of Physics, Saint~Petersburg State University, 198504 St.~Petersburg, Russia}
\def\affiSH{Physics and Astronomy School, University of Southampton, Highfield, Southampton, SO171BJ, UK}

\title{Multiple quantum beats of quantum confined exciton states in InGaAs/GaAs quantum well}
\author{A.~V.~Trifonov}
\author{I.~Ya.~Gerlovin}
\author{I.~V.~Ignatiev}
\author{I.~A.~Yugova}
\author{R.~V.~Cherbunin}
\affiliation{\affiSOLAB}
\author{Yu.~P.~Efimov}
\author{S.~A.~Eliseev}
\author{V.~V.~Petrov}
\affiliation{\affiRC}
\author{A.~V.~Kavokin}
\affiliation{\affiSOLAB} \affiliation{\affiSH}

\date{\today }

\begin{abstract}
Multiple quantum beats of a system of the coherently excited quantum confined exciton states in a high-quality heterostructure with a wide InGaAs/GaAs quantum well are experimentally detected by the spectrally resolved pump-probe method for the first time. The beat signal is observed as at positive as at negative delays between the pump and probe pulses. A theoretical model is developed, which allows one to attribute the QBs at negative delay to the four-wave mixing (FWM) signal detected at the non-standard direction. The beat signal is strongly enhanced by the interference of the FWM wave with the polarization created by the probe pulse. At positive delay, the QBs are due to the mutual interference of the quantum confined exciton states. Several QB frequencies are observed in the experiments, which coincide with the interlevel spacings in the exciton system. The decay time for QBs is of order of several picoseconds at both the positive and negative delays. They are close to the relaxation time of exciton population that allows one to consider the exciton depopulation as the main mechanism of the coherence relaxation in the system under study.
\end{abstract}

%\pacs{78.67.De, 78.55.Cr, 78.47.jg, 71.35.Cc}

\maketitle

\section*{Introduction}

Coherence of optical transitions between the ground and excited states of a quantum system is well explained as in classical as in quantum mechanical approaches. However, the mutual coherence of excited states is the purely quantum effect and cannot be treated classically. One of the brightest evidences of quantum coherence is the quantum beats (QBs)~\cite{Aleksandrov-1964,Dodd-1964, Haroche-PRL1973}. The QBs are the temporal oscillations of intensity of the emitted or absorbed light by a quantum system with several closely spaced quantum levels coherently excited by a short pulse. 

In semiconductors, the QBs are typically observed for discrete energy states of excitons interacting with light~\cite{Shah-USSSN1999}. The majority of experimental studies are devoted to QBs of the heavy- and light-hole excitons~\cite{Bartels-PRB1997,Lyssenko-PRP1993, Joschko-PRL1997,Mayer-PRB1995,Pal-PRB2003} as well as of the spin states of excitons and carriers~\cite{Bar-PRL1991,Amand-PRL1997,Gerlovin-PRB2002,Yugova-PRB2002,Kozin-PRB2002,Gerlovin-PRB2004,Shen-PRB2007,Gerlovin-PRB2007} . Attractive systems for the QBs observation are the quantum confined exciton states in quantum wells (QWs). Mutual coherency of these states is of particular interest because they are considered as promising systems for terahertz radiation~\cite{Planken-PRL1992,Dragoman-PQE2004,Kavokin-PRL2012} . Such quantum systems are also attractive because of possible beats of several, rather than two, quantum confined states. The QBs in multilevel system are theoretically discussed in Refs.~\cite{Leichtle-PRA1996, Eberly-PRL1980} . A multi-level system of the Landau level magneto-excitons has been experimentally studied in Ref.~\cite{Dani-PRB2008} but the QBs were observed only between the two lowest levels. No experimental studies of multi-frequency QBs are reported so far, to the best of our knowledge.

In this paper, we report on experimental study of time evolution of the coherently excited system of the quantum confined exciton states in a high-quality heterostructure with the InGaAs/GaAs QW. We have found that the excitation of several exciton states by a short laser pulse gives rise to the multi-frequency oscillations of reflectance of the heterostructure.  The theoretical analysis has shown that the oscillations are due to QBs of the coherently coupled exciton states. 

\begin{figure}[h]
\includegraphics[width=1\columnwidth]{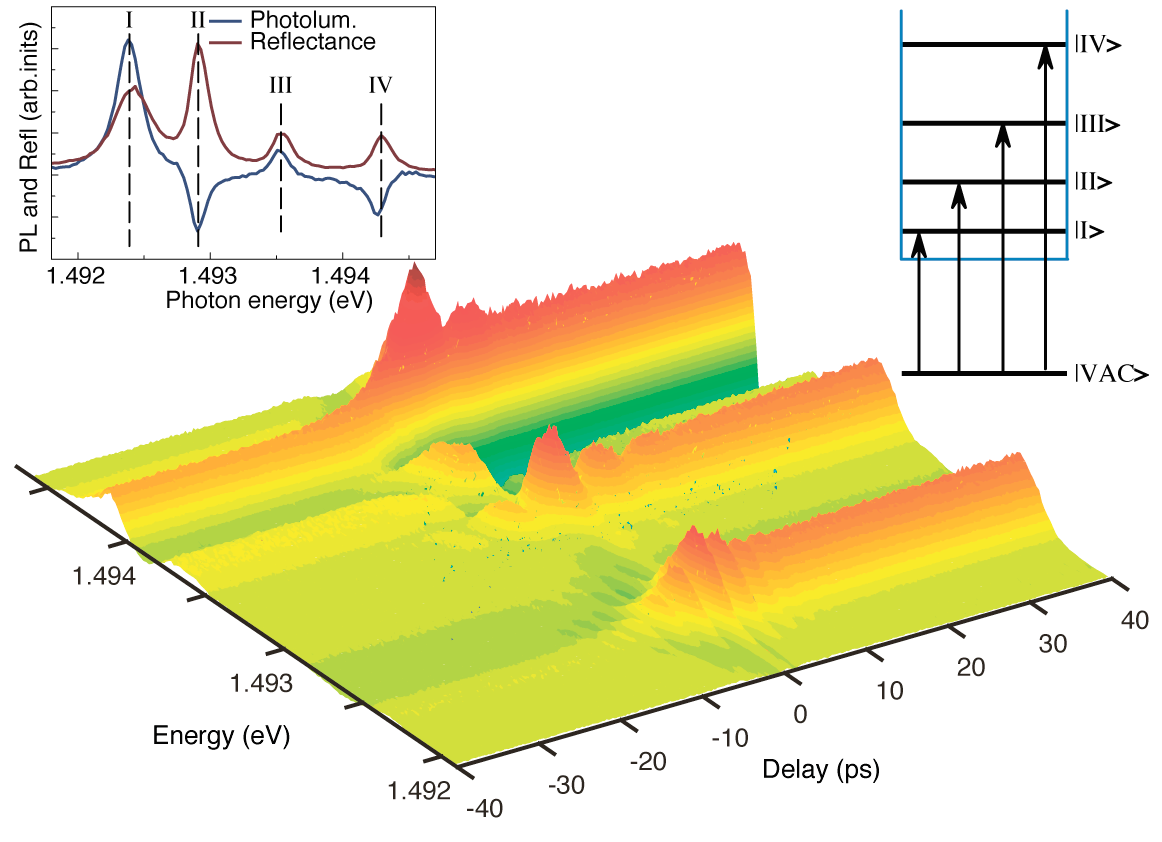}
\caption{(Color online) Spectrally resolved kinetics of pump-probe signals under coherent excitation of four exciton states. Left inset: photoluminescence (red) and reflectance (blue) spectra of the heterostructure with four exciton resonances (I - IV). Right insert: a simplified scheme of the exciton transitions.}
\label{fig1}
\end{figure}

We have studied a heterostructure with the 95-nm In$_{0.02}$Ga$_{0.98}$As/GaAs QW grown by the molecular beam epitaxy. The reflectance and photoluminescence spectra of the structure reveal a number of peculiarities related to the quantum confined exciton states in the QW (left inset in Fig.~\ref{fig1}). The detailed optical characterization of the structure is given in Ref.~\cite{Trifonov-PRB2015} . 

The kinetics of exciton states is studied by the pump-probe method, in which the modification of reflectance induced by a strong pump pulse is tested by a weaker probe pulse as a function of the time delay between the pump and probe pulses. The spectral width of the pump pulses was sufficient to coherently excite the four lowest quantum confined exciton states. The maximum of the pump beam spectrum was tuned to efficiently excite the weakest third and fourth exciton transitions. To measure the modulated reflectance at each particular exciton transition, the reflected probe beam was spectrally resolved in a 0.5-m spectrometer and detected by a photodiode connected with a lock-in amplifier and computer. 

Exciton dynamics for spectral range of four lowest exciton transitions is shown in Fig.~\ref{fig1}. The maximum of the pump-probe signal is observed for the fourth transition due to its efficient excitation by the pump beam. As seen from the figure, the amplitudes of spectral resonances depend on the delay by a complex manner, namely, there are rapidly decaying oscillations superimposed on the slowly varying background signal. The nature of slow component of the signal is related to the long-lived reservoir of non-radiative excitons as it is discussed in detail in Ref.~~\cite{Trifonov-PRB2015} . Here we will discuss the oscillating component, which we attribute to QBs. It is important to note that the QBs is observed as at positive delays as at negative ones when the probe pulse comes before the pump one.

\begin{figure}[h]
\includegraphics[width=1\columnwidth]{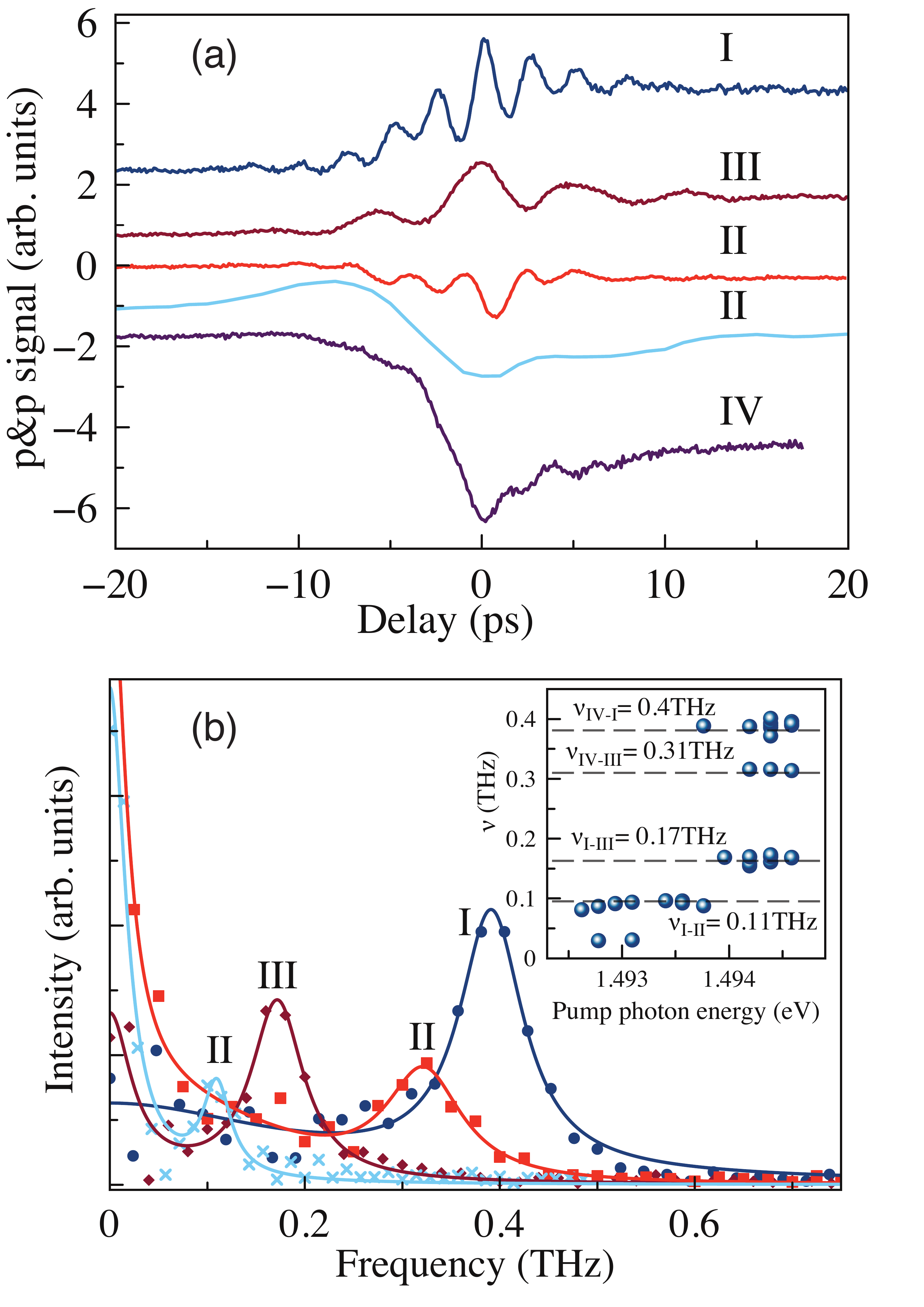}
\caption{(Color online) (a) Examples of reflectance kinetics detected at different spectrally selected exciton transitions noted near each curve. Curves I, II, III and IV are detected under predominant excitation of the state IV. Curve II${}^{*}$ was detected when only I and II states were excited. (b) Fourier analysis of the kinetics. Solid lines are the fits by Lorentzians. Notations I-II etc. indicate the frequencies of QBs of respective exciton states. Inset shows the frequencies of QBs obtained from the experiment (solid balls) and expected beat frequencies obtained from the energy spacing between the exciton states (dashed lines).}
\label{fig2}
\end{figure}

Figure~\ref{fig2} shows the reflectance kinetics for several exciton transitions as well as the Fourier analysis of the signals. As seen the frequencies of QBs obtained from the analysis for transitions to exciton levels I - III correspond to the energy spacing between respective exciton level and level IV (see inset in Fig.~\ref{fig2}). The oscillating component in the signal detected at the transition to level IV is relatively weak however the Fourier analysis allows one to reliably determine the QB frequency and to attribute it to the mutual coherence of the I and IV exciton states. When the pump beam coherently excites only the I and II exciton states, the QBs with smallest frequency, $\nu_{12}=0.094$~THz, is observed.

In a general case, analysis of the transient response of a multilevel quantum system, e.g., observed in the four-wave mixing or pump-probe signal, requires the identification of physical origin of temporal oscillations in the response. The oscillations can appear due to two different processes. The first one is related to the optical interference of polarizations created by the coherent excitation of the {\em independent} quantum systems. The oscillations are observed while the {\em polarization coherence} at optical frequencies is saved. The second process is the QBs of coherently excited states of the {\em single} quantum system. In this case, the polarization coherence is not required and the QBs are observed while the {\em mutual} coherence of excited states exists. 

Discrimination of these two processes in four-wave mixing experiments is a challenging problem~\cite{Koch-PRL1992,Erland-PRB1993} . On the contrary, pump-probe experiments allow one to easily identify the processes because oscillations in the pump-probe signal can arise only due to QBs of excited states. The interference of polarizations created by the pump beam in independent quantum systems is not detected in these experiments. The QBs can be detected because, when the mutual coherence of excited states is created, the probability of optical transition to the ground state of the quantum system is an oscillating in time function~\cite{Corney-PPS1964,Senitsky-PRA1977,Agarwal-PRA1977} . These oscillations of the probability give rise to the beating signal observed in kinetics of photoluminescence~\cite{Aleksandrov-1964,Dodd-1964} and of modulated reflectance~\cite{Pal-PRB2003} . By this reason, the oscillations observed in our experiments can be definitely treated as QBs of quantum confined exciton states. 

Theoretical analysis shows that, when the coherent superposition of several exciton states is prepared, the electric field of light wave of the reflected probe beam consists of several component for each particular exciton transition $k$:
\begin{equation}
\label{Eq1}
E_k \sim \sum \limits_j E_{kj} e^{[ i (\varepsilon_j-\varepsilon_k)\tau/ \hbar ]}
\end{equation}
[see Eq. (7) in Supplementary Materials]. Each component $j$ contains an complex exponential factor, which is the oscillating function of time delay, $\tau$, between the pump and probe pulses. The oscillation frequency, $\omega_{kj} = (\varepsilon_k - \varepsilon_j)/\hbar$ is determined by the energy spacing between the state $k$, which is detected in the experiment, and one of other coherently excited exciton states. According to Eq.~ (7) in [Supplementary Materials], the contribution of each component depends on the spectrum of exciting pulses controlling the efficiency of excitation of respective state. Indeed, when the IV-th quantum confined exciton state has been predominantly excited in the experiment (see Fig.~\ref{fig2}), the QB frequencies obtained under detection of pump-probe signal at the exciton transitions I - III were determined by the energy difference between respective exciton state and state IV. At the same time, the QBs obtained at the IV-th exciton transition are mainly determined by the interference of states IV and I, because the optical transition from state I obeys the maximal oscillator strength~\cite{Trifonov-PRB2015} .

The above discussion was focused on QBs observed at the positive delays when the probe pulse comes to the sample after the pump pulse. However, as it is seen in Fig.~\ref{fig2}(a), the QBs are observed as a bright effect also at the negative delays when the probe pulse precedes the pump one. This phenomenon requires separate discussion. One of possible mechanisms suggested in Ref.~\cite{Hawkins-PRB2003} could be as follows. The probe pulse creates an oscillating polarization due to coherent excitation of several exciton states. The strong enough pump pulse coming with some delay may destroy the coherence because of creation of excitons and free carriers. This coherence breaking gives rise to the step-like decrease of the secondary exciton emission in the direction of reflected probe beam, which results in the oscillating in time delay signal~\cite{Senitsky-PRA1977} However, as an analysis shows, oscillations in the detected signal should be weak in this case.

We suggest another origin of the oscillating signal at the negative delays, namely the diffraction of the exciton polarization created by the pump pulse on the population grating created by the joint action of the probe and pump pulses. This is the four-wave mixing (FWM) signal detected in the direction of the reflected probe beam. We should mention that the standard direction for the FWM detection is determined by $2k_1 - k_2$, where $k_1$ and $k_2$ are the projections of wave vectors of the pump and probe light waves on the QW plane, respectively~\cite{Shah-USSSN1999}. We have used the nonstandard direction (determined by $k_2$) for the FWM detection, which has got an important advantage as it is discussed below. 

\begin{figure}[h]
\includegraphics[width=1\columnwidth]{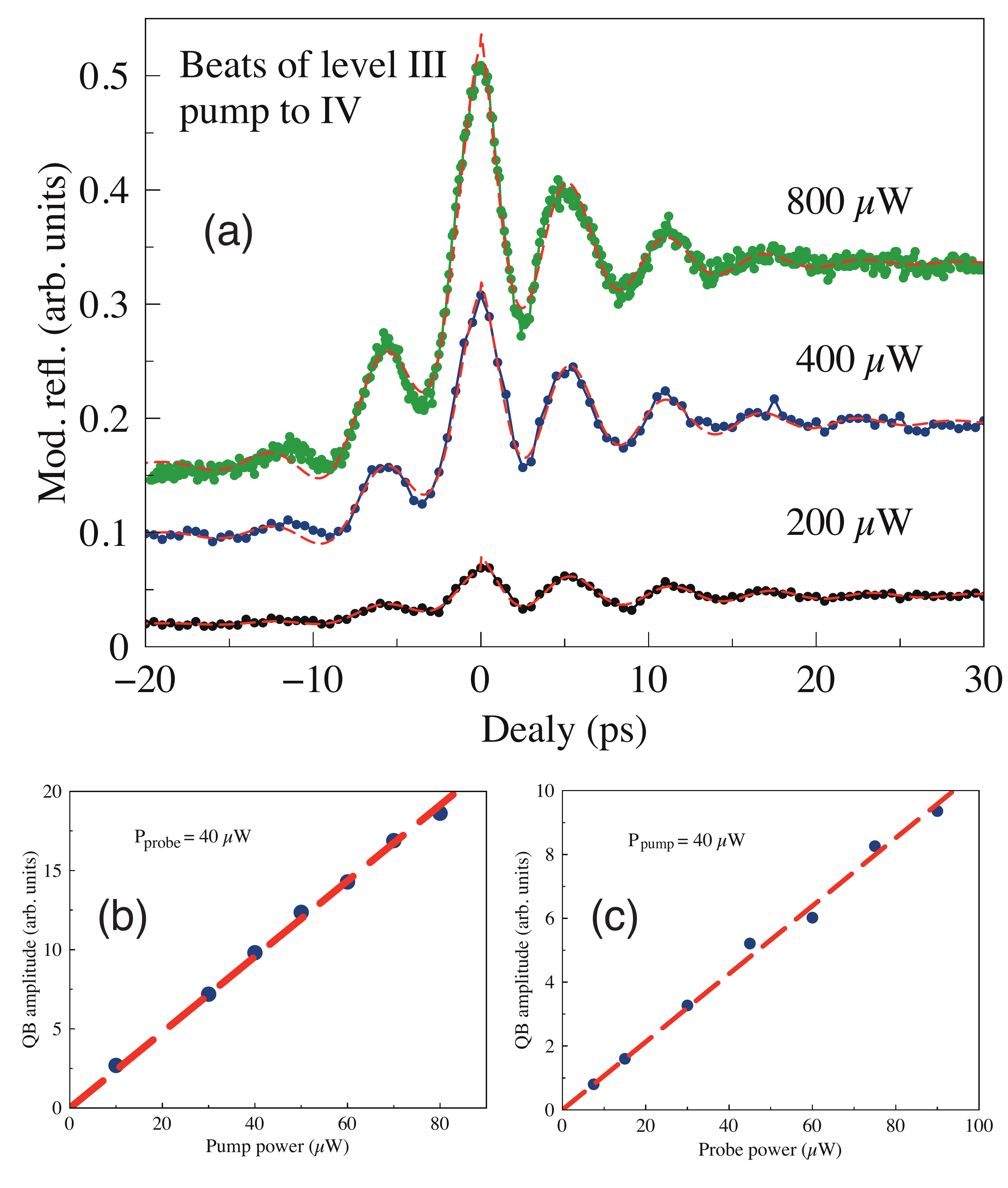}
\caption{(Color online)(a) Pump-probe signals measured at the III-rd exciton transition at different pump powers given near each curve. $P_{probe} = 40 \mu$W. The observed QBs are due to the interference of the III-rd and IV-th exciton states. Dashed lines are the fits by function $Q(t) = Q_0 \cos(\omega t) \exp(-|t|/\tau)$ superimposed on a smooth function describing the background signal. (b) and (c) show the dependences of beat amplitude, $Q_0$, on the pump and probe powers, respectively.}
\label{fig3}
\end{figure}

To verify the above assumption, we have studied the dependence of detected signal on the pump and probe powers. For the FWM detected in the standard direction, this signal should linearly depend on intensity of the pulse coming first and should be square-dependent on the intensity of the delayed pulse~\cite{Shah-USSSN1999} . At first glance, it would be expected similar behavior of the FWM detected by us in the nonstandard direction. Experiment, however, demonstrates linear dependence of the FWM signal on both the pump and probe powers at their variation at least within one order of magnitude (Fig.~\ref{fig3}). Further increase of the intensities gives rise to rapid decay of QBs, most probably due to accumulation of the non-radiative excitons~\cite{Trifonov-PRB2015} , so that the beat amplitude cannot be reliable determined. 

To understand the obtained seeming contradiction, we should take into account one more light wave propagating in this nonstandard direction. This is the secondary emission of the excitons excited by the probe beam. It is a relatively intense wave, which interferes with the weaker FWM wave so that the detected signal is proportional to $I_{pp} |E_p|^2 |E_{pr}|^2$[see Supplementary Materials].  

So the interference with the reflected probe wave explains the linear dependences observed experimentally.

The experimentally obtained time evolution of the QBs allows us to directly estimate the lifetime of mutual coherence of the quantum confined exciton states. For this purpose we fit the oscillating component of the delay dependences by function $Q(t) = Q_0 \cos(\omega t) \exp(-|t|/\tau)$. Thus obtained decay constants are: $\tau_1 = 3.5\pm1$~ps for the I-st exciton state and $\tau_n = 7\pm2$~ps for other states. These values are well agree with those obtained by a Fourier analysis from the line broadening studied in Ref.~~\cite{Trifonov-PRB2015} . The decay constant for each exciton transition is found to be the same for positive and negative delays within an experimental error. 

The latter result is nontrivial because, as it is discussed above, the signals observed at positive and negative delays are formed by different processes. The QB decay in the pump-probe signal (positive delay) is controlled by the relaxation of the {\em mutual} coherence of excitonic states whereas the decay in the FWM signal (negative delay) is due to the relaxation of {\em polarization} coherence at the optical frequency. Typically, the optical frequency dephasing is the fastest process in a quantum system controlled by interaction with different quasiparticles (nonradiative excitons, carriers, phonons) as well as by the inhomogeneous broadening of the quantum ensemble~\cite{Shah-USSSN1999} . At the same time, the mutual coherence of exciton states should survive much longer and degrade with the population relaxation time.

Almost total coincidence of decay times for positive and negative delays points out that, for the structure under study, the major contribution to the decay of mutual (quantum) coherence and of coherence of exciton transitions comes from the same processes. Since the QB decay times are close to the exciton depopulation time, $\tau \approx 6.5$~ps, reported in Ref.~\cite{Trifonov-PRB2015} for this heterstructure, we may conclude that the main process of the decoherence of exciton states is the rapid population relaxation. Such rapid depopulation is mainly due to very high rate of radiative exciton recombination.

In conclusion, the experimental study of exciton dynamics in a high-quality heterosctructure with a wide InGaAs/GaAs QW allowed us to detect a new type of QBs, which are due to the quantum interference of several quantum confined exciton states in the QW. The beat signal is detected as at positive delays as at negative ones. A theory of QBs of the coherently excited multiple quantum states is developed. The theoretical analysis shows that the QBs at negative delays are observed due to the FWM effect detected in the nonstandard direction of the reflected probe beam. The amplitude of the FWM signal is strongly enhanced by the interference of the FWM light wave with the strong polarization wave created by the probe pulse. At the positive delays, the QBs are detected in the ordinary pump-probe signal. Surprisingly, the decay time of QBs detected at the positive and negative delay is the same within an experimental error although the relaxation mechanisms are seems to be different. At positive delays, this is the relaxation of mutual coherence of the exciton states while the relaxation at negative delays is caused by the dephasing of optical waves. The origin of this unusual behavior is the high quality of heterostructure under study, in which the main mechanism of the coherence relaxation is related to the depopulation of exciton states due to radiative recombination and scattering of radiative and nonraditive excitons. 

\section*{Acknowledgments}

The authors thank M. M. Glazov for fruitful discussions.
Financial support from the Russian Ministry of Science and Education
(contract no. 11.G34.31.0067), SPbU (grants No. 11.38.213.2014), RFBR 
and DFG in the frame of Project ICRC TRR 160 is acknowledged. 
The authors also thank the SPbU Resource Center ``Nanophotonics'' 
(www.photon.spbu.ru) for the sample studied in present work. 
A.~V.~Kavokin acknowledges the support from EPSRC Established Career Fellowship.

\end{document}